%% file: tot.tex
\documentclass{emulateapj}
        
\shorttitle{INTEGRAL/IBIS CENSUS OF THE SKY BEYOND 100 keV}
\shortauthors{A.~Bazzano}

\begin{document}

\title{INTEGRAL/IBIS CENSUS OF THE SKY BEYOND 100 keV}

\author{A.~Bazzano\altaffilmark{1}, J.B.~Stephen\altaffilmark{2}, M.~Fiocchi\altaffilmark{1},
 A.J.~Bird\altaffilmark{3}, L.~Bassani\altaffilmark{2}, A.J.~Dean\altaffilmark{3}, A.~Malizia\altaffilmark{2},
P.~Ubertini\altaffilmark{1}, F.~Lebrun\altaffilmark{4}, R.~Walter\altaffilmark{5} and C.~Winkler\altaffilmark{6} }

\altaffiltext{1}{IASF-Roma/INAF, via Fosso del Cavaliere 100,
I-00133 Roma, Italy; angela.bazzano@iasf-roma.inaf.it}
\altaffiltext{2}{IASF-Bologna/INAF, Via Gobetti 101, I-40126 Bologna,
Italy;} \altaffiltext{3}{School of Physics and Astronomy,
University of Southampton, Southampton S017 1BJ,
UK;}\altaffiltext{4}{CEA-Saclay,DAPNIA/Service d'Astrophysique, F91191 Gif-sur-Yvette Cedex, France;}
\altaffiltext{5}{Geneva Observatory, INTEGRAL SDC
, Chemin d'Ecologia 16, 1291 Versoix,
Switzerland;}\altaffiltext{6}{ESA-ESTEC, RSSD,
Keplerlaan 1, 2201 ZA Noordwijk, Netherlands;}

\begin{abstract}

We report on the first census of INTEGRAL/IBIS detections
($\gtrsim 4\sigma$ significance) above 100 keV based on the Core
Program and public Open time observations up to April 2005. There
are 49 sources detected in the 100-150 keV band of which 14 are also seen 
in the 150-300 keV range. The low energy sample is dominated by X-ray binary
systems of both low and high mass, but also includes 10 active
galaxies. Of the binary systems that are detected above 150 keV,
more than 50\% are associated with black hole candidates, often
reported as microquasars. The present survey results are then used
to construct LogN-LogS curves for galactic and extragalactic
objects in the 100-150 keV band: above a 1 mCrab sensitivity limit
we expect that around 200 galactic sources and almost 350 active
galaxies populate the sky above 100 keV. While the contribution of
individual point sources to the total Galactic emission has 
been estimated to be around 70-80$\%$ between 100-300
keV, we find that active galaxies detected above 1 mCrab
account for only about
3$\%$ of the cosmic hard X-ray background in the 100-150 keV band.
\end{abstract}


\section{Introduction}
One of the primary goals of the INTEGRAL observatory
\citep{wink03}, and in particular of the IBIS imager
\citep{uber03}, is to provide a survey of the high
energy sky ($>$20 keV). To this end a number of teams have made an
effort to provide a systematic analysis of the data sets, as
available to them at the time, so as to provide the general
community with consolidated source information: so far 2 general
surveys \citep{bird04, bird06}, several surveys of specific areas of
the sky \citep{revn04a,revn06,molk04,lebr04} and a few AGN
catalogues \citep{beck06,bass06} have been published. A
common factor throughout all of these activities is that the source
search concentrated on the 18-100 keV regime where the IBIS
sensitivity to point sources is optimal. Emission at higher
energies has not been specifically investigated for or
characterised in terms of source typology and emission mechanism
except for a few objects. In this letter we focus our search above
100 keV and report the first census of the sky above this energy;
as a result we provide for the first time the LogN-LogS
relationship for galactic and extragalactic sources in the 100-150
keV band as well as an estimate of the galactic source
contribution to the total Galactic emission and of the AGNs to the
cosmic hard X-ray background.

\section{Pre-INTEGRAL soft $\gamma$-ray sky}

The first ever attempt to survey the sky at high energy (26-1200
keV) was performed with the Sky Survey Instrument on Ariel V \citep{coe82}, which
provided the first, and to date only, galactic LogN-LogS relation above 100 keV. 
This was followed by the HEAO 1-A4 survey in the 40-180 keV band \citep{levi84}. 
Of the 44 sources detected, 14 had a
formal statistical significance $\ge 4\sigma$ in the 80-180 keV
band, i.e. well above the sensitivity limit of 36 mCrab. The 14 objects detected 
comprise 6 low mass X-ray binaries
(LMXB), 4 high mass X-ray binaries (HMXB), 3 active galactic
nuclei (AGN) and the Crab. Two of the LMXB sources were at that
time associated with A1742-294, near the Galactic Centre, and GX
5-1. Subsequent observations by SIGMA/GRANAT
detected 2 persistent black hole candidates (BHC) close to these
sources that were probably responsible for the detected high
energy tails. Assuming that these new identifications are correct,
we can categorize the HEAO-A4 binary compact objects as 6 BHCs, 2
neutron stars (NS) and 2 objects without a clear identification so
far. A further step in our knowledge of the high energy sky was
provided by the SIGMA sky survey \citep{revn04b} which
contains 37 sources of which only 6 were detected in the 100-200
keV band (5 black holes and the Crab). 

A further insight into the high energy regime was provided by a
catalogue of 309 sources
amassed from the literature \citep{maco99}. Most (189)
of these were gamma ray emitters, detected only
above 1 MeV. Of the
remainder, only 50 were reported to emit above 100 keV
and these include several X-ray Novae (e.g. Nova Muscae, GRO
0422+32) sporadically detected only when flaring once or twice per
decade.
The high sky coverage of the GRO-BATSE experiment provided a list of  
around 50 objects in the 70-160 keV band \citep{harm04}. The list is
dominated by X-ray binaries both of low and high mass; however a
few AGN are also reported together with supernova remnant/pulsar
associations. At higher
energies ($E>700$ keV), the  GRO/Comptel survey \citep{scho00}
lists 32 persistent  sources including various types of galactic
objects and 9 AGN. The galactic sources are divided into
spin-down pulsars (7), Cygnus X-1, the Nova GROJ0422+32, 3
unidentified EGRET objects and the Crab. The extragalactic sources
are all Blazars except for CenA. This survey also
reported sky regions where MeV emission features are observed.
These detections, when compared to those in the hard X-ray domain,
clearly indicate that there is a transition region where accreting
mechanics are not longer dominant (except in the brightest
sources) while synchrotron processes become important and
nucleosynthesis starts taking place. This transition region is
still largely unexplored despite its potential importance; this
paper is an attempt to start filling in this observational gap.

\section{SPI/INTEGRAL}
Analysis of the first year of SPI data \citep{bouc05} resulted
in a list of 63 hard X-ray point sources detected in the 25-50 keV
band during the Galactic Plane Survey and Galactic Center Deep
Exposure; only 20 sources were detected in the 50-150 keV band
(above $4\sigma$) and 4 sources (above $5\sigma$) in the 150-300
keV band. One main objective of this study was to provide a direct
estimate of the source contribution to the diffuse galactic
emission, which was estimated to be around 90\% at 100 keV while
above 250 keV, diffuse electron-positronium emission dominates
over the discrete source component. This result is in substantial
agreement with IBIS/ISGRI data which indicate that in the 20 to 200
keV range the known binary sources account for 86$\%$ to 74$\%$ of
the total galactic emission \citep{lebr04}.

\section{IBIS dataset and analysis}
The data used herein belong to the Core Program and public Open
Time observations and span from revolution 46 (February 2003) to
revolution 309 (April 2005) inclusive; this represents a
significant extension both in exposure time and sky coverage with
respect to the second IBIS catalogue \citep{bird06} with more than
4000 extra pointings ($\sim$8Ms) being analysed. A detailed
description of the data analysis and source extraction criteria
can be found in the above reference, the only difference 
being the use of an updated version (4.2) of the standard OSA
software.
For this search  we have used the 100-150 and 150-300 keV flux
maps and have adopted a $4\sigma$ significance threshold level for
the source extraction; this is lower than the $5\sigma$ level used in 
the general source search but is justified by the {\em a posteriori} 
confirmation that all detected sources have a low energy counterpart.
Sources detected only occasionally,
e.g. in one or two revolutions only, are not considered in the
present sample. Staring data (which are noisier than dithering
observations), early exposures performed with the instrument
set-up not finalized and poor quality data (including solar
flares, telemetry gaps, etc.), were not included in the present
mosaic.
For each excess, the flux extracted from the 100-150 and 150-300
keV light curves was then used to estimate the source strength
with the use of a normalization factor extracted from the Crab
calibration data-set. The light curves were used for the flux
extraction as they provide a more reliable result than the mosaic,
and also provide a confirmation of the source detection. Once a
list of reliable excesses was produced, we proceeded to identify
the sources responsible for the emission. We find that all
excesses have a counterpart at lower energies and all objects
reported in this survey were already detected below 100 keV.

\section{The high energy IBIS sky}

\begin{figure}
\plotone{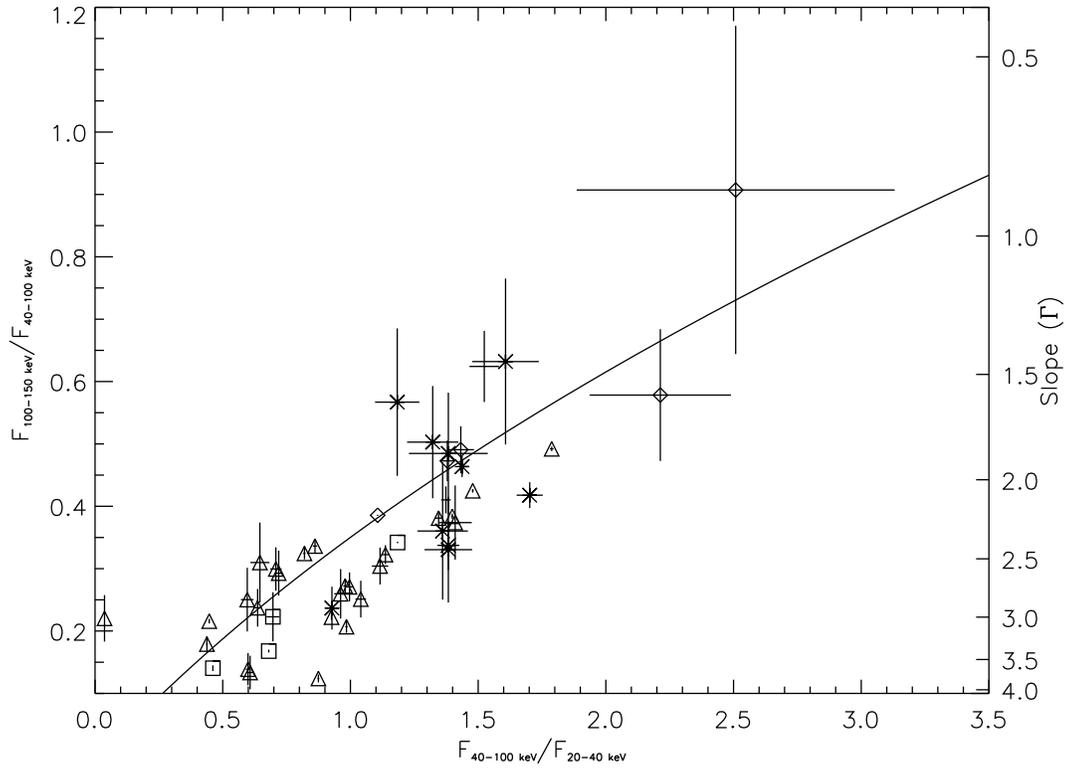} \caption{The High energy hardness ratio (100-150 keV/40-100 keV) 
as a function of the low energy hardness ratio (40-100 keV/20-40 keV). 
The stars are AGN, diamonds are PSR, triangles and squares are LMXB and HMXB and 
no symbol are unidentified. The curve is the locus of ratios as a function of photon slope (left-hand axis)}
\label{hardness}
\end{figure}

In Table \ref{tab1} we list the 49
sources seen in the 100-150 keV band of which 14 are also detected in
the 150-300 keV range. The table provides the key parameters such
as the source name, type and average fluxes for the 20-40, 40-100,
100-150 and 150-300 keV band in mCrab. In the two (highest) bands
of interest, 1 mCrab corresponds to 3.2 and 4.8$\times10^{-12}$
erg cm$^{-2}$ s$^{-1}$ respectively. The 2 lower bands are given
so as to have a consistent set of data updated with respect to
that reported in \citet{bird06}. Also, BeppoSAX detections above
100 keV are reported in order to
provide a flux comparison. The 49 sources listed have also been
checked against studies reporting specific and more detailed
analyses and the associated references are also provided. 
We find substantial agreement between our fluxes and
those previously reported taking into account expected source
variability. 
In our high energy catalogue we have 35
accreting galactic objects (26 LMXBs, 6 HMXBs, 2 Anomalous X-Ray
Pulsar (AXPs) and 1 Soft $\gamma$-Ray Repeater (SGR)), 3 isolated
pulsars, 10 AGNs and 1 unidentified source suggested to be a
possible BHC in a LMXB \citep{cap06}. It is interesting to note
that while in the 100-150 keV band various source types are
detected, the higher band is totally dominated by BHCs.
In Fig. \ref{hardness}, we plot the hardness ratios, HR2 versus HR1, where HR1
is defined as the ratio of the 40-100 keV to the 20-40 keV flux
and HR2 as the ratio of the 100-150 keV to the 40-100 keV flux.
It is evident from the figure that there is a range of hardness
ratios. The two hardest sources (upper right corner) are anomalous
X-ray pulsars while the softest object is Sco X-1.

\section{The Source Distribution and logN-logS}

Even though the number of sources in this sample is limited, it is
still useful to construct the logN-logS number flux relationship
particularly as the sky above 100 keV is still virtually
unexplored. With so few sources, it serves more as a tool with
which to estimate the total number of objects which will become visible as the
survey becomes more complete than as an instrument with which to
study the spatial distribution. Apart from the paucity of sources,
there are other complications involved in forming this
relationship. The sensitivity limit in any direction depends on
the exposure at that point, and is extremely non-uniform due to
the pointing strategy of INTEGRAL \citep{bird06}. Furthermore, for
the case of galactic sources, there is a strong relationship
between the intrinsic distribution of the sources and the
exposure, while the source list consists of an ensemble of several
distinct populations (LMXBs, HMXBs etc.) each of which have
different logN-logS parameters \citep{dean05}.

\begin{figure}
\plotone{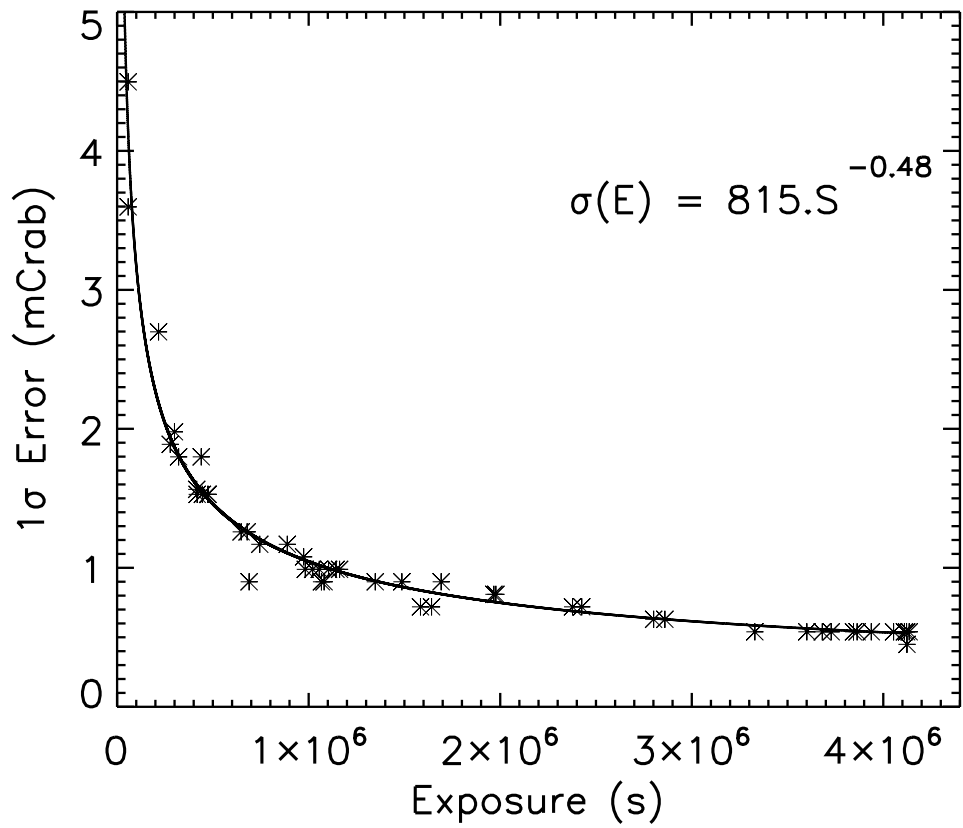} \caption{The error on the source fluxes
as a function of source exposure. This relationship is used to
compute the flux limit for the entire exposure map.}
\label{limitingFlux}
\end{figure}

\begin{figure}
\plotone{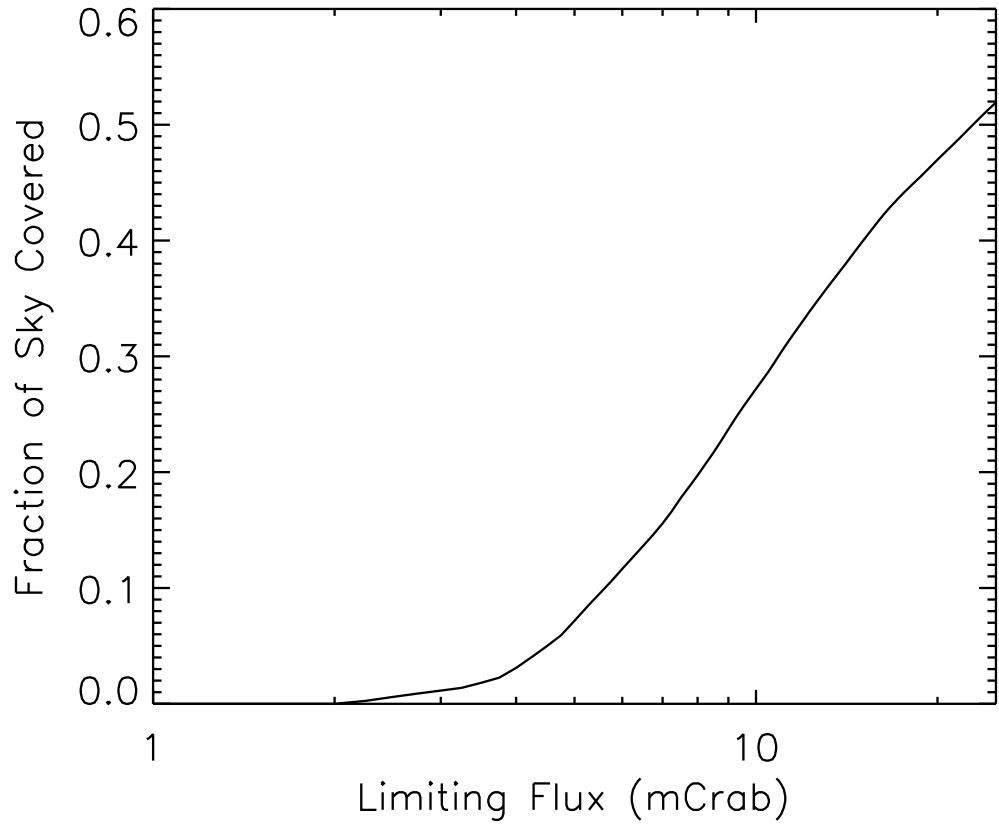} \caption{The all-sky fractional coverage as a function of ($4\sigma$) limiting flux.} \label{coverage}
\end{figure}

\begin{figure}
\plotone{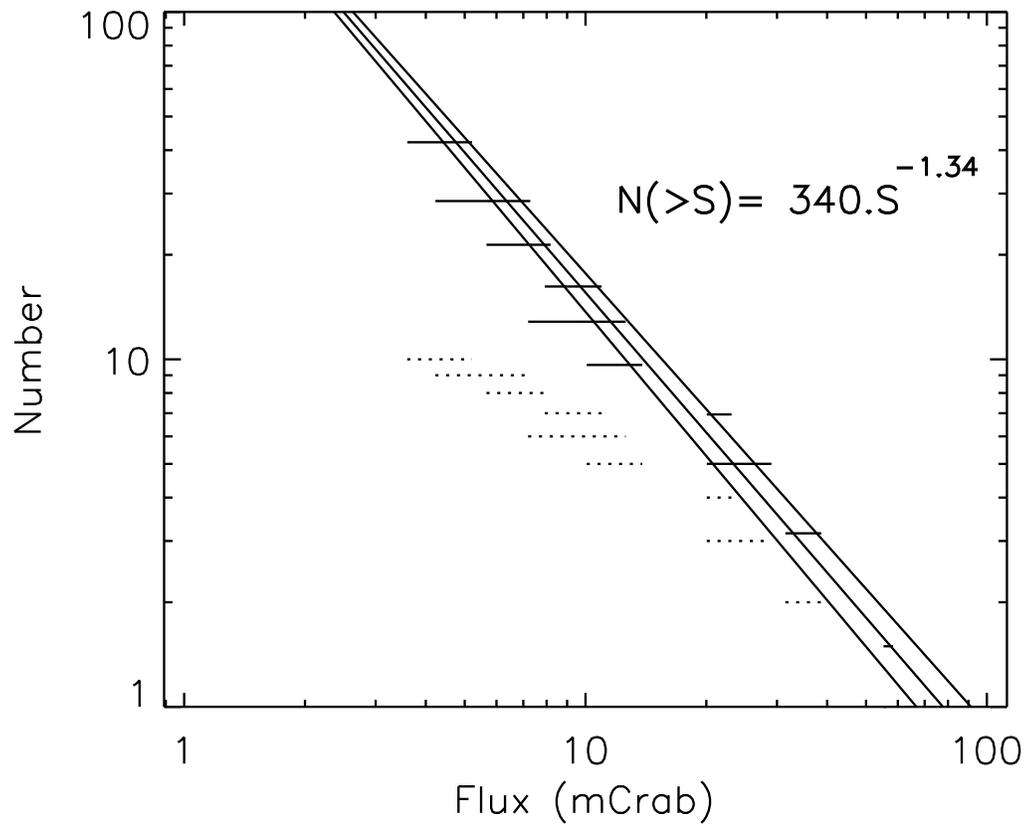} \caption{100-150 keV full sky number-flux relationship for the 10
AGN in our sample. Data points for both before (lower, dotted) and
after correction for exposure are shown as are the best-fit power
law with $1\sigma$ limits.} \label{flux}
\end{figure}

\begin{figure}
\plotone{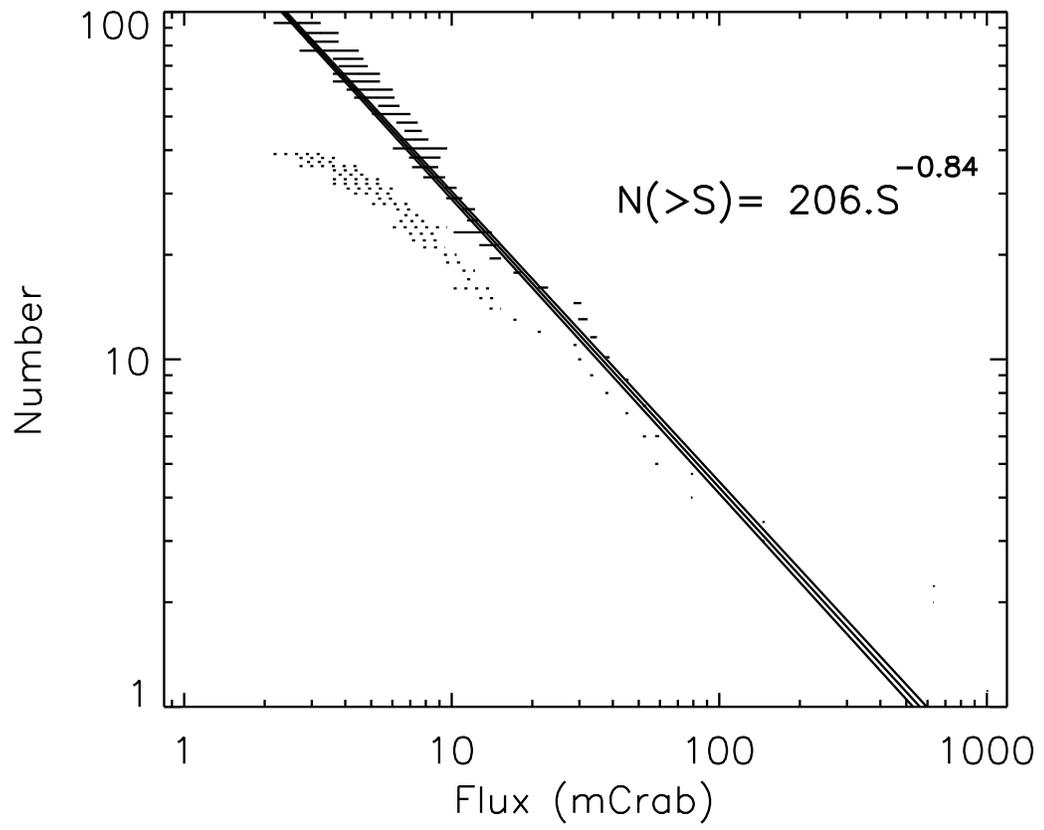} \caption{As Fig. \ref{flux}, but for the 39
galactic sources in the sample.} \label{galactic}
\end{figure}

Given these caveats, we first convert exposure to $1\sigma$ limiting
sensitivity using the relationship shown in Fig.
\ref{limitingFlux}, between the flux
errors and effective exposure for the sources used in this study. From this we obtain the 
sky coverage as a function of $4\sigma$ limiting flux as shown in Fig. \ref{coverage}.
We then create the `raw' logN-logS for both the 10 AGNs and the 39
Galactic sources independently. These distributions are 
corrected for the exposure limit and sky coverage using two
methods. The first is model independent (or, rather, assuming an
isotropic underlying distribution), while the second assumes a
model distribution based on the observed sources.
Fig. \ref{flux} shows the results for the AGN. Both the `raw' data
and the corrected values normalised to full sky coverage are shown
where the horizontal bars indicate the errors on the flux of the
$N^{th}$ source. Unsurprisingly, for the AGN we find that the best
model and the isotropic cases are identical (i.e. the best model
is an isotropic distribution). The best-fit power law relationship
and the ($1\sigma$) statistical limits are indicated by the solid
lines and are given by $N(>S)=(340\pm12)S^{-1.34\pm0.04}$. The
slope is slightly flatter than the expected value of 1.5, but we
can cross-check the number by making reference to the HEAO1-A4
results for the 80-180 keV all-sky survey. The
limiting flux of this survey was 36 mCrab, and from our
relationship we would expect to see $3\pm1$ AGN above that flux
over the whole sky, in agreement with the 3 listed in the A4
catalogue (3C273, NGC4151 and CenA).
In Fig. \ref{galactic} the results for the 39 galactic sources are
shown. The best-fit relationship is given by
$N(>S)=(206\pm3)S^{-0.84\pm0.01}$. The quoted errors are purely
statistical, but some idea of the systematic error can be obtained
by comparing this fit with that assuming {\it a-priori} an
isotropic source distribution. This leads to
$N(>S)=(206\pm23)S^{-0.84\pm0.06}$. Once again this can be tested
by comparing the predictions for a full-sky survey at the
sensitivity of the HEAO-A4 instrument. At 36 mCrab we should
expect to see $10\pm2$ galactic sources while there are 11 in the
A4 catalogue. The slope is very similar to that obtained for the
LMXB sources in the 20-100 keV band from the first IBIS catalogue
\citep{dean05}, which is to be expected as 26 of the 39
galactic sources in this sample are LMXB's.

\section{Conclusions}
A further use of the present catalogue is to estimate the
contribution of galactic point sources to the total Galactic
emission and that of AGN to the cosmic diffuse background in the
energy bands explored in the present paper. The total galactic
point source contribution has been estimated by summing the
100-150 keV fluxes from all sources detected within the central
radius excluding all
AGNs and Sco-X1 which is located at high galactic latitude. The
flux obtained amounts to $2.17\pm0.01\times10^{-4}$ and
$4.29\pm0.10\times10^{-5}$ ph cm$^{-2}$s$^{-1}$keV for
100-150 and 150-300 keV respectively; this implies a contribution
of 83\% and 68\% in the two bands in agreement with previous
and current INTEGRAL results \citep{bouc05,lebr04}.
The percentage contribution does not increase much
as we go to fainter fluxes as sources a factor of 100 weaker will
bring the point source component to 85\% contribution. The above
estimates further confirm previous indications that point sources
largely dominate the total Galactic emission. In a similar way,
the extragalactic logN-logS distribution derived from the present
survey  provides a total AGN emissivity above 1 mCrab of $\sim
0.003$ keV cm$^{-2}$ keV $^{-1}$ s$^{-1}$ sr$^{-1}$ in the  100-150
keV or approximately 3\% of the total intensity of the
extragalactic hard X-ray background in the same energy band
\citep{grub99}. Extrapolation of our LogN-LogS by a factor
of 100 towards fainter fluxes will account for 15\% of the
extragalactic background above 100 keV; this value increases to 20\% 
if we adopt a LogN-LogS slope of 1.5.
In short we just start to
uncover the source populations responsible for the hard X-ray
background. As the INTEGRAL mission lifetime is extended (so far
confirmed till 2010), more sky will be surveyed and more exposure
will be accumulated resulting in an increase in the number of
objects detected. At a
limiting flux of 1 mCrab (which is likely reachable in the entire
lifetime of the mission) around 200 galactic objects may be
detected and more than 300 AGN identified at high energy.

\acknowledgements
This research was supported by the Italian Space Agency under contract 
I/023/05

\input{tab_tot}
\end{document}

%% file: tab_tot.tex
\newcommand{\gtsimeq}{\raisebox{-0.6ex}{$\,\stackrel
        {\raisebox{-.2ex}{$\textstyle >$}}{\sim}\,$}}
\begin{table}
\center\caption{IBIS detections in (100-150)keV and (150-300)keV
energy band}
\scriptsize
\begin{tabular}{lcccccccc}
\hline
 name& class&IBIS&IBIS&IBIS&IBIS&SAX&SAX&ref\\
& & (20-40)&(40-100)&(100-150)&(150-300)&(100-150)&(150-200)&\\
&&keV&keV&keV&keV&keV&keV&\\
& &mCrab&mCrab&mCrab&mCrab&mCrab&mCrab&\\
\hline
4U0142$+$61 & AXP& $1.5\pm0.3$ & $3.4\pm0.5$ & $8\pm2$ &    &  & &(1)\\
Crab & PWN PSR & $1000.0\pm0.4$ & $1000.0\pm0.6$ & $1000\pm2$ & $1000\pm6$& $1000\pm3$$^{(c)}$&$1000\pm3$ &\\
MKN3 & AGN & $4.2\pm0.2$ & $6.1\pm0.4$ & $10\pm2$ & & & &\\
VelaPulsar & PWN PSR & $6.74\pm0.09$ & $8.4\pm0.1$ & $10.3\pm0.7$ & & & &\\
GS0836-429 & LMXB T B &$30.58\pm0.09$ & $27.2\pm0.1$ & $14.6\pm0.7$ & & &  &\\
NGC4151 &AGN Sey1.5 &$32.0\pm0.6$ & $40\pm1$ & $35\pm4$ & &$28.5^{+0.3}_{-0.3}$$^{(b)}$&$18.0^{+0.3}_{-0.5}$&(2)\\
NGC4388 &AGN Sy1 & $15.9\pm0.7$ & $17\pm1$ & $25\pm5$ & & & &(3)\\
3C273 & AGN QSO & $7.7\pm0.3$ & $9.2\pm0.6$ & $12\pm2$ & & $23.7\pm0.7$$^{(a)}$ &$27\pm1$ &(4)\\
NGC4507 &AGN Sy1h & $8.7\pm0.4$ & $10.7\pm0.6$ & $10\pm3$ &&& &\\
NGC4945 &AGN Sey2 & $13.2\pm0.3$ & $20.3\pm0.4$ & $22\pm1$ & &  $23.1^{+0.6}_{-0.6}$$^{(b)}$ & &\\
CenA &AGN Sey2 &$36.5\pm0.2$ & $47.4\pm0.4$ & $57\pm2$ & $65\pm6$&   $49.5^{+0.3}_{-0.3}$$^{(b)}$&$52.0^{+0.5}_{-1.0}$&(5)\\
Cir galaxy &AGN Sy1h &$13.6\pm0.2$ & $11.4\pm0.3$ & $7\pm1$ &&$3.4^{+1.1}_{-0.5}$$^{(c)}$ & &\\
PSRB1509-58 &PSR &$8.5\pm0.2$ & $11.0\pm0.3$ & $14\pm1$ & && &(6)\\
XTEJ1550-564 &LMXB T BH & $70.8\pm0.2$ & $114.4\pm0.3$ & $146\pm1$ & $155\pm4$&  $115^{+1}_{-1}$$^{(a)}$&$114.2^{+0.5}_{-1.5}$&(7)\\
4U1608-522 &LMXB T B A  & $14.3\pm0.2$ & $7.7\pm0.3$ & $5\pm1$ & & & &\\
ScoX-1 &LMXB Z  & $629.0\pm0.4$ & $21.0\pm0.6$ & $12\pm2$ & & & &\\
4U1630-47&LMXB T BHC & $44.2\pm0.2$ & $34.4\pm0.3$ & $30\pm1$ & $30\pm4$& $48.5^{+0.3}_{-0.3}$$^{(a)}$&$43.0^{+0.5}_{-0.5}$&(8)\\
4U1636-536&LMXB B A & $22.6\pm0.2$ & $13.0\pm0.3$ & $8\pm1$ & & &&\\
OAO1657-415&HMXB XP & $78.6\pm0.2$ & $42.5\pm0.3$ & $7.3\pm0.9$ &  &  & &\\
GX339-4&LMXB T BH &$28.3\pm0.2$ & $34.4\pm0.3$ & $34\pm1$ & $31\pm4$& $43.4^{+0.3}_{-0.3}$$^{(b)}$ & $28.8^{+0.5}_{-0.5}$    &(9)\\
4U1700-377&HMXB &$196.21\pm0.09$ & $120.6\pm0.1$ & $52.5\pm0.7$ & $23\pm3$& & &(10)\\
4U1702-429&LMXB B A & $16.1\pm0.2$ & $10.3\pm0.3$ & $8.0\pm0.9$ & &$10.2^{+0.3}_{-0.3}$$^{(b)}$&&\\
4U1705-440&LMXB B A & $23.7\pm0.2$ & $13.0\pm0.3$ & $4.5\pm0.9$ & &$14.2^{+0.3}_{-0.3}$$^{(b)}$& &\\
IGRJ17091-3624&BHC?& $8.78\pm0.09$ & $10.9\pm0.1$ & $11.6\pm0.6$ & & & &(11)\\
XTEJ1720-318&LMXB T BHC &$2.59\pm0.09$ & $3.3\pm0.1$ & $3.2\pm0.5$ & & & &(12)\\
GRS1724-30&LMXB B A G&$19.21\pm0.09$ & $17.0\pm0.1$ & $12.0\pm0.5$ &&$26.1^{+0.6}_{-0.6}$$^{(b)}$& &\\
GX354-0&LMXB B A&$39.17\pm0.09$ & $15.5\pm0.1$ & $7.2\pm0.5$ & & & &\\
GX1+4&LMXB XP&$38.98\pm0.09$ & $30.8\pm0.1$ & $9.9\pm0.5$ &&$10.5^{+0.3}_{-0.3}$$^{(b)}$&&\\
SLX1735-269 & LMXB B &$9.33\pm0.09$ & $8.4\pm0.1$ & $5.9\pm0.5$&&&&\\
1E1740.7-294 & LMXB BHC&$35.84\pm0.09$ & $45.3\pm0.1$ & $45.1\pm0.5$ & $37\pm2$& & &(13)\\
KS1741-291 & LMXB T B&$5.64\pm0.09$ & $4.9\pm0.1$ & $3.3\pm0.5$ &&& &\\
A1742-294 & LMXB T B&$13.86\pm0.09$ & $7.5\pm0.1$ & $2.7\pm0.5$ &&& &\\
IGRJ17464-3213 &  LMXB BHC & $28.08\pm0.09$ & $20.8\pm0.1$ & $17.5\pm0.5$ & $20\pm2$&& &(14)\\
SLX1744-299 & LMXB B & $8.31\pm0.09$ & $5.4\pm0.1$ & $4.1\pm0.5$ & & & &\\
IGRJ17597-2201 & LMXB B D & $7.02\pm0.09$ & $6.6\pm0.1$ & $4.3\pm0.5$ & & & &\\
GRS1758-258 & LMXB BHC & $53.58\pm0.09$ & $71.6\pm0.1$ & $78.9\pm0.5$ & $70\pm2$& & &(15)\\
SGRJ1806-20 & SGR & $3.05\pm0.09$ & $4.2\pm0.1$ & $6.8\pm0.6$ & & $2.4\pm0.3$$^{(a)}$&$2.2\pm0.5$  &(16)\\
4U1812-12 & LMXB B A &  $25.6\pm0.2$ & $26.3\pm0.3$ & $22\pm1$ & $18\pm4$& & &(17)\\
GS1826-24 & LMXB B&$79.26\pm0.09$ & $66.3\pm0.1$ & $38.2\pm0.7$ & $12\pm3$& $32.2^{+0.3}_{-0.3}$$^{(b)}$&&\\
PKS1830-211& AGN QSO & $2.8\pm0.2$ & $3.5\pm0.3$ & $4.4\pm0.8$ & & & &(18)\\
1RX1832-33 & LMXB B T G & $10.81\pm0.09$ & $10.9\pm0.3$ & $8.6\pm0.8$ &  &  & &\\
KES73 & SNR AXP &$2.0\pm0.2$ & $4.0\pm0.3$ & $6\pm1$ & &$21\pm2$$^{(a)}$&$31\pm3$&\\
4U1909+07 & HMXB T XP& $14.3\pm0.2$ & $9.0\pm0.3$ & $5.2\pm0.9$ &  & &  &\\
AqlX1 & LMXB B A T &$9.6\pm0.2$ & $5.6\pm0.3$ & $4.5\pm0.9$ & &$3.6^{+1.1}_{-0.8}$$^{(a)}$&$3.3^{+1.0}_{-0.6}$&(19)\\
GRS1915+105& LMXB BH & $260.8\pm0.2$ & $105.4\pm0.3$ & $59\pm1$ & $47\pm4$& $82.6^{+0.2}_{-0.2}$$^{(b)}$&$55.5^{+0.2}_{-0.1}$    &(20)\\
CygX-1 & HMXB  BH &$665.7\pm0.3$ & $712.5\pm0.4$ & $632\pm2$ & $522\pm6$&$950^{+3}_{-3}$$^{(b)}$&$870^{+5}_{-5}$&(21)\\
EXO2030+375 & HMXB XPBe T & $38.8\pm0.2$ & $20.4\pm0.4$ & $3.6\pm0.9$&&& &(22)\\
CygX-3 & HMXB &$204.1\pm0.2$ & $85.1\pm0.3$ & $31\pm1$ & $20\pm5$&  $18.0^{+0.3}_{-0.3}$$^{(b)}$&$12.1^{+0.5}_{-0.5}$&(23)\\
IGRJ21247+5058 & AGN Sy~1? & $5.6\pm0.2$ & $7.0\pm0.4$ & $6.0\pm1.5$ & & & &\\
\hline
\end{tabular}
\tablecomments{ \scriptsize Class are according with Bird et al.
2006. Beppo-Sax fluxes in millicrab measured by PDS in the band
100-150 and 150-200 keV respectivily. Only sources detected with
signal to noise $>$ 4 are reported. 1mCrab=3.2$\times10^{-12} erg$
$cm^{-2} s^{-1}$ in the 100-150 keV range.
1mCrab=4.8$\times10^{-12} erg$ $cm^{-2} s^{-1}$ in the 150-300 keV
range.\\The notes indicate the model used to fit Beppo-Sax data:
(a) simple power law, (b) cut-off power law and (c) broken power
law. }
 \label{tab1}
 \tablerefs{\scriptsize(1) Kuiper et al.
2006; (2) Beckmann et al. 2005; (3) Beckmann et al. 2004; (4) Courvoisier et al. 2003; (5)
Rotschild et al. 2005; (6) Shaw et al. (7) Sturner et al. 2005;
(8) Tomsick et al., 2005 (9) Belloni et al. 2006; (10) Orr et al.
2004; (11) Capitanio et al. 2005; (12) Cadolle-Bel et al. 2004;
(13) Del Santo et al. 2005; (14) Capitanio et al. 2004; (15) 
Pottschmit et al. 2006; (16) Mereghetti
et al. 2005; (17) Tarana et al. 2006; (18) De Rosa et al., 2005;
(19) Molkov et al. 2004; (20) 
Hannikainen et al. 2005; (21)
Cadolle-Bel et al. 2006; (22) Camero
Arranz, A. et al. 2005 (23) Vilhu et al. 2003}
\end{table}